\pdfoutput=1
\documentclass[11pt,a4paper]{article}

\usepackage[T1]{fontenc}
\usepackage[utf8]{inputenc}
\usepackage{lmodern}
\usepackage{amsmath,amssymb}
\usepackage{graphicx}
\usepackage{booktabs}
\usepackage{siunitx}
\usepackage{cite}
\usepackage[colorlinks=true,
            allcolors=blue]{hyperref}

\graphicspath{{figures/}}

\usepackage[margin=2.5cm]{geometry}

\title{Encapsulated macroscopic WS$_2$ monolayers enable room-temperature exciton-polariton lattices}

\author{
Shiyu Huang$^{1,4,5,\dagger,*}$,
Sander Scheel$^{1,4,5,\dagger}$,
Jiang Qu$^{2,4,\dagger}$,
Johannes Düreth$^{1,4}$,
\\
Dominik Horneber$^{1,4,5}$,
Edith Wietek$^{3,4}$,
Simon Widmann$^{1,4}$,
Libo Ma$^{2,4}$,
\\
Monika Emmerling$^{1,4}$,
Martin Kamp$^{1,4}$,
Simon Betzold$^{1,4}$,
\\
Alexey Chernikov$^{3,4}$,
Sven Höfling$^{1,4}$,
and Sebastian Klembt$^{1,4,5}$
\\[1.5ex]
\small $^{1}$Julius-Maximilians-Universität Würzburg, Physikalisches Institut,\\
\small  Lehrstuhl für Technische Physik, Am Hubland, 97074 Würzburg, Germany
\\
\small $^{2}$Leibniz Institute for Solid State and Materials Research
(IFW Dresden),\\ \small Helmholtzstraße 20, 01069 Dresden, Germany
\\
\small $^{3}$Institute of Applied Physics,\\
\small Technische Universität Dresden, 01062 Dresden, Germany
\\
\small $^{4}$Würzburg-Dresden Cluster of Excellence ctd.qmat,\\
\small Würzburg-Dresden, Germany
\\
\small $^{5}$Current affiliation:
\small Julius-Maximilians-Universität Würzburg, Physikalisches Institut,\\
\small  Lehrstuhl für Experimentelle Physik 1, Am Hubland, 97074 Würzburg, Germany\\
\small 
\small $^{\dagger}$These authors contributed equally to this work.
\\
\small $^{*}$Corresponding author:
\href{mailto:shiyu.huang@uni-wuerzburg.de}
{shiyu.huang@uni-wuerzburg.de}
}

\begin{document}

\title{Encapsulated macroscopic WS$_2$ monolayers enable room-temperature exciton-polariton lattices}
\maketitle

\begin{abstract}
Large-area, optically homogeneous monolayer semiconductors are a critical prerequisite for scalable room-temperature polaritonics and for realizing polariton lattices extending across many unit cells. Yet, the small size, optical inhomogeneity, and device-to-device variability of conventional exfoliated flakes have remained major obstacles. Here, we overcome these limitations using 1-dodecanol-encapsulated WS$_2$ monolayers that combine millimeter-scale coverage with remarkably uniform optical properties over lateral distances approaching \SI{300}{\micro\meter}. Integrated into a tunable open microcavity, these monolayers exhibit robust room-temperature exciton--photon strong coupling, evidenced by a pronounced anti-crossing and a Rabi splitting of $\hbar\Omega_{\mathrm{R}} \approx \SI{31}{meV}$. Leveraging the exceptional uniformity of this platform, we realize a two-dimensional polaritonic kagome lattice and directly resolve its characteristic band structure. Angle-resolved spectroscopy reveals Dirac dispersive bands together with a weakly dispersive flat-band-like branch within the $s$-band, in good agreement with a linear non-interacting model. Complementary momentum- and real-space imaging further identifies the associated bond-centered and site-centered mode profiles. These results establish large-area WS$_2$ monolayers in open microcavities as a scalable platform for engineering polariton band structures and exploring synthetic quantum materials under ambient conditions.
\end{abstract}

% =========================================================
%\section{Introduction}
% =========================================================

%Introduction
Artificial lattices for photons, excitons and hybrid light--matter quasiparticles provide a powerful route to engineer band structures and emulate condensed-matter Hamiltonians, including flat-band physics and topological band concepts~\cite{Jo2012UCatom-Kagome,Tang2020Photonic,Polini2013Artificial}. Exciton--polaritons, formed by strong coupling between cavity photons and excitons, combine a low effective mass with optical nonlinearities, enabling access to collective driven--dissipative phenomena~\cite{PhysRevLett.69.3314,Deng2010Exciton}.

Two-dimensional semiconductors, particularly transition-metal dichalcogenide (TMDC) monolayers, are an attractive materials platform for room-temperature polaritonics and related applications. Monolayers such as MoS$_2$ and WS$_2$ have direct bandgaps and tightly bound excitons that remain stable at room temperature~\cite{Mak2010direct-gap,Splendiani2010,Wang2018RevModPhys,Zhao2013_ACSNano}, and the exciton binding energy in WS$_2$ reaches many hundreds of meV, which is about an order of magnitude larger than in conventional semiconductors~\cite{Green2013_SiExciton,Miller1985_GaAs,Monemar1996_GaN_Exciton}. These strong optical resonances enable exciting directions for fundamental research and technology, including exciton-mediated superconductivity~\cite{Kavokin2016_ExcitonSC,VonMilczewski2024_ExcitonSC_TMD}, localized single-photon emitters~\cite{Gao2023_SinglePhoton}, polaritons in layered two-dimensional materials~\cite{Schneider_TMDCPolaritonsReview2018} and TMDC-based biosensing~\cite{Perkins2013}. In microcavities and photonic structures, TMDC monolayers feature prominently in room-temperature polariton condensates~\cite{Zhao2021}, WS$_2$-based Tamm-plasmon polaritons~\cite{Yang2025}, polaritons in engineered photonic potentials~\cite{Lackner2021_WS2-ML-SC} and switching-controlled polariton screening in MoS$_2$ microcavities~\cite{Mondal2025}; however, such platforms are severely limited by micron-scale exfoliated flakes and individual devices, where limited lateral size and device-to-device variability hinder reproducible and systematic studies as well as the realization of polariton lattices spanning many unit cells. Large-area and optically uniform monolayers are therefore essential for implementing complex photonic band structures, accessing technologically relevant regimes of exciton--polaritons, and exploiting the flexibility of emerging photonic architectures. 

A second key requirement is tunability of the microcavity. Although monolithic sputtered-DBR microcavities provide, in principle, a complementary and potentially scalable route to strong-coupling systems~\cite{Federolf2026}, key parameters such as cavity length and exciton--photon detuning remain fixed after fabrication. This substantially limits the accessible polaritonic band structures and the flexibility needed to explore detuning-dependent regimes. Open optical microcavities address this limitation by enabling real-time tuning of the mirror separation, providing \textit{in situ} control of the detuning and polariton composition, while allowing for integration with TMDC monolayers and photonic lattices~\cite{Dufferwiel2015,Lackner2021_WS2-ML-SC}. 
Controlling the spatial gradient of the cavity photon energy has recently been used to engineer photonic potential landscapes for WS$_2$ polariton propagation~\cite{Yang2026_ACSPhotonics}. The open geometry also allows controlled alignment between a TMDC monolayer on the bottom mirror and a patterned top mirror, enabling different lattice geometries to be explored by exchanging patterned mirrors while keeping the same active monolayer.

Gold-assisted exfoliation overcomes the size limitations of conventional exfoliation by yielding high-quality, large-area monolayers~\cite{Science_Gold-tape-exfoliation2020}. Here, we use 1-dodecanol as an encapsulation layer for WS$_2$ as an alternative to conventional hBN~\cite{PhysRevX.7.021026,Li_Dodecanol-WS2_2023}. The resulting fully 1-dodecanol-encapsulated WS$_2$ monolayers (D/WS$_2$/D) combine millimeter-scale coverage with enhanced and spatially homogeneous optical properties. Integrated into a tunable open microcavity, they enable room-temperature strong coupling and the realization of a two-dimensional polaritonic kagome lattice, providing a large-area route to engineered polariton band structures.

In the following, we establish the optical uniformity of D/WS$_2$/D monolayers over hundreds of micrometers, demonstrate room-temperature strong coupling in a planar open cavity, and realize a WS$_2$ polaritonic kagome lattice whose band structure is resolved by angle-resolved tomography and captured by a linear Gross-Pitaevskii model.

% =========================================================
\section*{Results}
% =========================================================

\subsection*{Macroscopic WS$_2$ monolayers with uniform optical properties}
\begin{figure}[t]
\hspace*{\fill}
\includegraphics[width=1\textwidth]{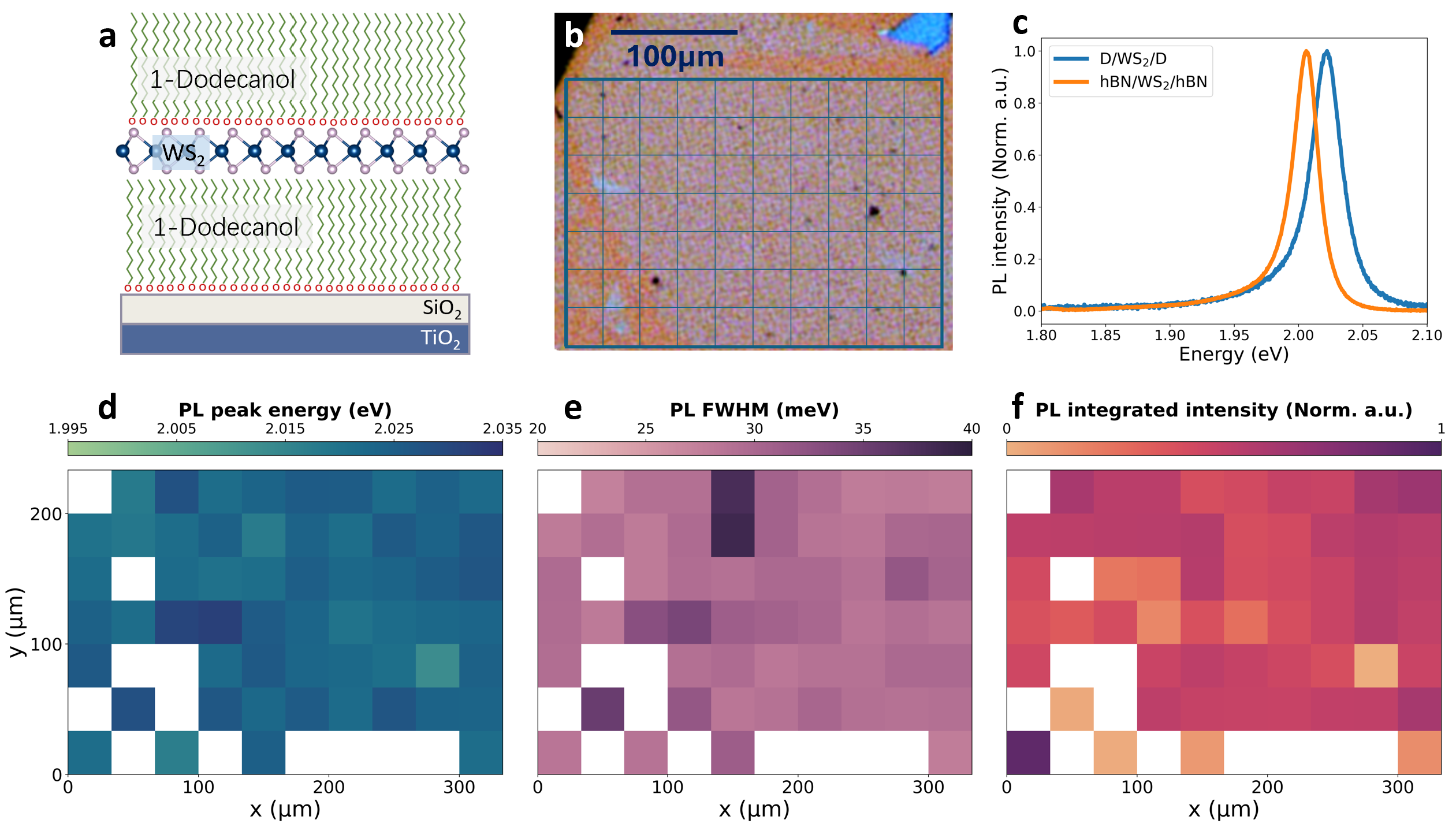}
\hspace*{\fill}
\caption{\textbf{Optical uniformity of 1-dodecanol-encapsulated WS$_2$ monolayers.}
\textbf{a}, Schematic illustration of a WS$_2$ monolayer with both bottom and top 1-dodecanol encapsulation (D/WS$_2$/D).
\textbf{b}, Optical microscope image of D/WS$_2$/D with a $\SI{329}{\micro m}\times\SI{230}{\micro m}$ field of view divided into a $10 \times 7$ grid; the light-blue area highlights the WS$_2$ monolayer. Regions with stronger blue contrast correspond to multilayer or bulk WS$_2$, whereas areas without blue contrast indicate no WS$_2$ coverage. The blue shaded rectangle marks the region used for PL mapping in panels \textbf{d}--\textbf{f}.
\textbf{c}, Representative room-temperature PL spectra measured under identical excitation and collection conditions for a D/WS$_2$/D monolayer and an hBN-encapsulated WS$_2$ reference flake.
\textbf{d}--\textbf{f}, Spatial maps of the exciton PL properties extracted from Lorentzian fits: peak energy (\textbf{d}), linewidth (\textbf{e}) and integrated intensity (\textbf{f}); white pixels correspond to positions outside the monolayer region (multilayer/bulk or no WS$_2$ coverage) and are excluded from the monolayer analysis.}
\label{fig: Uniformity}
\end{figure}

We use gold-tape-assisted exfoliation (Supplementary Fig.~S1) to fabricate WS$_2$ monolayers with millimeter-scale coverage and continuous domains extending over hundreds of micrometers, enabling spatially resolved characterization over large areas, which is essential for reproducible polariton measurements over hundreds of lattice sites. An overview of the encapsulated monolayer architecture and the mapped region is shown in Fig.~\ref{fig: Uniformity}a,b.

Maintaining optical quality over such length scales requires effective encapsulation. However, applying hBN to macroscopic TMDC monolayers is challenging, motivating alternative approaches that retain key benefits such as reduced contamination and a controlled dielectric environment~\cite{PhysRevX.7.021026,Raja2019}. Li \textit{et al.} showed convincingly that 1-dodecanol provides such an alternative for WS$_2$, leading to a substantially improved optical response~\cite{Li_Dodecanol-WS2_2023}. Following this approach, we encapsulate our gold-exfoliated WS$_2$ monolayers with 1-dodecanol, which forms a self-assembled molecular layer and yields enhanced and more uniform optical properties across the monolayer.

To quantify optical uniformity, we performed spatially resolved photoluminescence (PL) mapping under $\SI{532}{nm}$ continuous-wave excitation. A focused laser spot was scanned over a $10\times7$ grid, and PL spectra were recorded at each position. Figure~\ref{fig: Uniformity}c shows room-temperature PL spectra for the D/WS$_2$/D monolayer and an hBN-encapsulated WS$_2$ reference measured under identical conditions; optical microscope images of the two cavity-integrated samples are shown in Supplementary Fig.~S2.
The spectra were fitted with a Lorentzian lineshape, commonly used for excitonic PL in monolayer WS$_2$ at room temperature~\cite{Fang2022,Selig2016,PhysRevX.7.021026}. From the fits, we extract the peak energy, full width at half maximum (FWHM) and integrated PL intensity, yielding the maps shown in Fig.~\ref{fig: Uniformity}d--f for an $\SI{329}{\micro\meter}\times\SI{230}{\micro\meter}$ region.

The fitted peak energy shows a standard deviation of $\SI{3.2}{meV}$ and a coefficient of variation (standard deviation/mean) of CV = $\SI{0.16}{\percent}$. This spread is considerably smaller than the mean PL linewidth of $\gamma_{\mathrm{PL}}\sim\SI{30}{meV}$; accordingly, the quoted variation reflects modest shifts within a broadened line and also includes fit-related uncertainty.

At room temperature, the PL linewidth is primarily broadened by exciton--phonon scattering, with additional contributions from inhomogeneity due to disorder, strain and charge fluctuations. We measure a mean linewidth of $\gamma_{\mathrm{PL}}=\SI{29.8}{meV}$, which lies at the low end of the typical $\SIrange{25}{50}{meV}$ range reported for monolayer WS$_2$ at room temperature~\cite{Selig2016,PhysRevX.7.021026,Lackner2021_WS2-ML-SC}. Engineering of the dielectric and photonic environment can reduce disorder-related broadening and yield some of the narrowest room-temperature linewidths reported for WS$_2$~\cite{Fang2022}, but exciton--phonon interactions remain the dominant contribution at ambient conditions. The linewidth is spatially uniform in our maps (cf. Fig.~\ref{fig: Uniformity}e), with a standard deviation of $\SI{2.18}{meV}$ and a CV = $\SI{7.31}{\percent}$.

The integrated PL intensity shows the largest relative variation (CV = \SI{35.53}{\percent}), consistent with its sensitivity to localized non-radiative centers. Nevertheless, the intensity remains comparatively uniform across a lateral length scale of \SI{300}{\micro\meter}, indicating spatially uniform recombination response. Together, these results establish 1-dodecanol-encapsulated WS$_2$ monolayers as a suitable large-area platform for macroscopic photonic and polaritonic devices.

\subsection*{Room-temperature exciton-polaritons in a tunable open cavity}
The open cavity is a tunable optical resonator formed by two distributed Bragg reflectors (DBRs), each comprising nine TiO$_2$/SiO$_2$ pairs with individual layer thicknesses of $d = \lambda/(4n_i)$, separated by an adjustable air gap that can be reduced below \SI{2}{\micro\meter}. The bottom DBR carrying the D/WS$_2$/D monolayer (on a SiO$_2$ spacer near a cavity-field antinode) is mounted on piezoelectric nanopositioners that provide three-dimensional positioning and tilt control (up to \SI{10}{\degree}). The top DBR (planar or patterned) is positioned with nanopositioners for cavity-length tuning and lateral alignment. The planar cavity design, including the calculated field profile, reflectivity spectrum and mesa geometry, is summarized in Supplementary Fig.~S4.

\begin{figure}[htbp!]\hspace*{\fill}\includegraphics[width=1.0\textwidth]{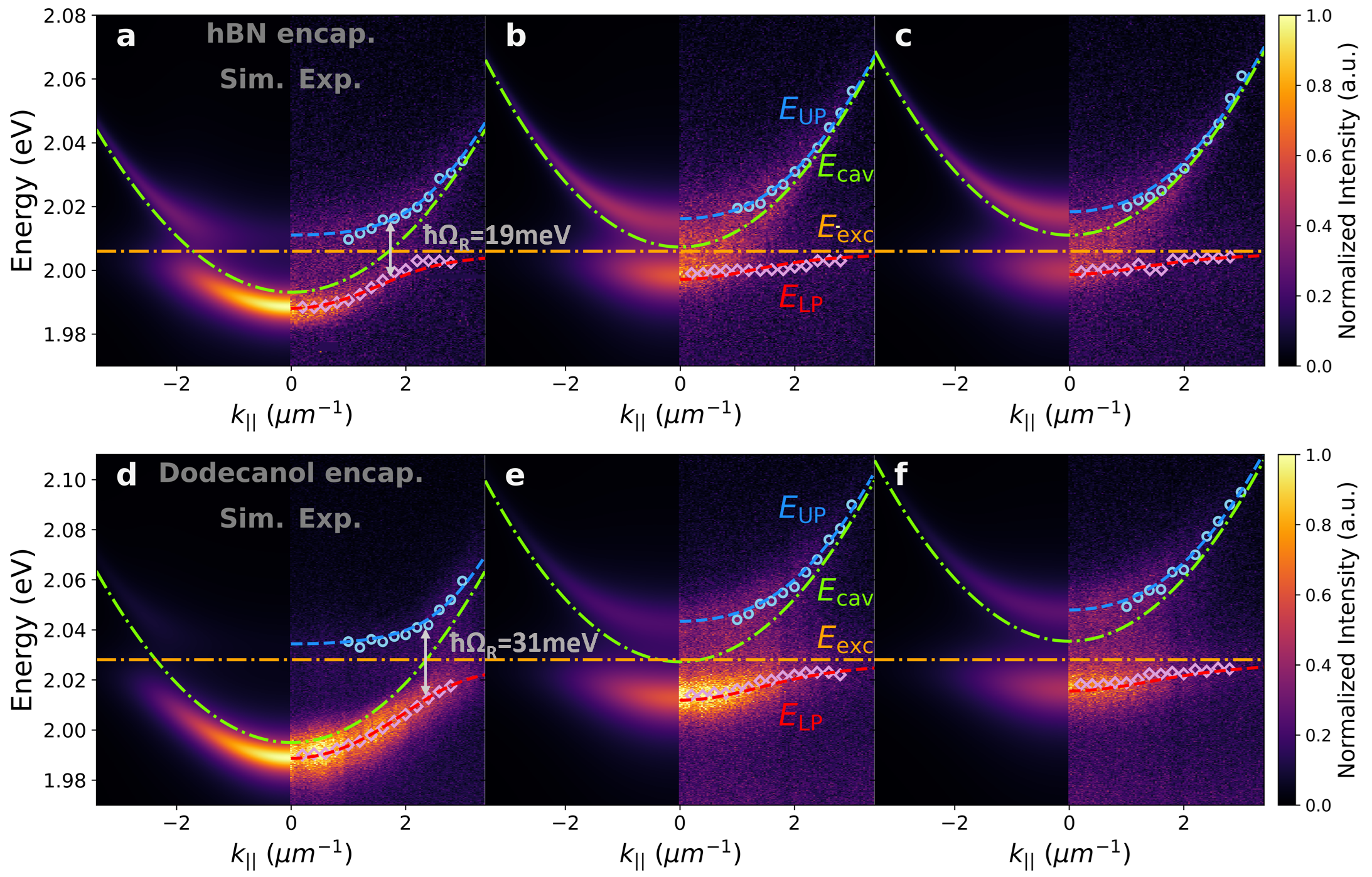}\hspace*{\fill}\caption{\textbf{Planar-cavity strong coupling and Boltzmann-population modelling for hBN- and 1-dodecanol-encapsulated WS$_2$ monolayers.}\textbf{a--c}, hBN-encapsulated WS$_2$ reference sample at detunings $\Delta=\SI{-12.91}{\meV}$ (\textbf{a}), $\SI{1.28}{\meV}$ (\textbf{b}) and $\SI{5.00}{\meV}$ (\textbf{c}).\textbf{d--f}, 1-dodecanol-encapsulated D/WS$_2$/D sample at detunings $\Delta=\SI{-33.09}{\meV}$ (\textbf{d}), $\SI{-0.83}{\meV}$ (\textbf{e}) and $\SI{7.37}{\meV}$ (\textbf{f}). In each panel, the left half shows the calculated momentum-resolved PL intensity from a Boltzmann-population model evaluated at $T=\SI{300}{\kelvin}$, and the right half shows the corresponding experimental angle-resolved PL spectrum. Open circles mark the extracted polariton peak positions. Blue and red dashed lines are the fitted upper and lower polariton branches ($E_{\mathrm{UP}}$ and $E_{\mathrm{LP}}$). Green and orange dash-dotted lines indicate the uncoupled cavity and exciton dispersions ($E_{\mathrm{cav}}$ and $E_{\mathrm{exc}}$).}\label{fig: planar-SC}\end{figure}

To benchmark strong coupling in our platform and to compare encapsulation strategies under identical measurement conditions, we record angle-resolved PL spectra at room temperature while tuning the cavity length. The evolution of the longitudinal cavity modes with cavity length is shown in Supplementary Fig.~S5. In both cases, the same planar top mirror is used, and the bottom mirrors have the same DBR design. Figure~\ref{fig: planar-SC} summarizes PL dispersions for an hBN-encapsulated WS$_2$ reference flake (Fig.~\ref{fig: planar-SC}a--c) and for the large-area D/WS$_2$/D monolayer (Fig.~\ref{fig: planar-SC}d--f). In both samples, tuning the cavity mode through the exciton produces well-defined upper and lower polariton branches that display a pronounced anti-crossing. At negative detuning the lower polariton is predominantly cavity-like and follows the bare cavity dispersion, whereas towards zero detuning it shifts towards the exciton energy and flattens around $k_{\parallel}=0$, consistent with an increasing excitonic fraction and an enhanced effective mass. A full detuning series is shown in Supplementary Fig.~S6.

The peak positions of the upper and lower polariton branches are extracted from the momentum-resolved PL spectra and fitted with a coupled-oscillator model for a single cavity mode coupled to the exciton resonance. The resulting fits reproduce the dispersions and their detuning dependence accurately (Fig.~\ref{fig: planar-SC}). From the near-resonant splitting, we obtain a Rabi splitting of $\hbar\Omega_{\mathrm{R}}^{\mathrm{hBN}}=\SI[separate-uncertainty]{19.0 \pm 1.2}{\meV}$ for the hBN-encapsulated reference and $\hbar\Omega_{\mathrm{R}}^{\mathrm{D}}=\SI[separate-uncertainty]{31 \pm 0.5}{\meV}$ for the D/WS$_2$/D sample (see Fig.~\ref{fig: planar-SC}a--c and d--f, respectively).

To place the observation of a normal-mode splitting on a quantitative footing, we compare the splitting to the relevant linewidths using the standard strong-coupling criterion for an exciton coupled to a single cavity mode (see Refs.~\cite{Savona1995,Savona1997,Keeling2020}), 

\begin{equation}
    \Omega_{\mathrm{R}} =2g > \frac{\gamma_{\mathrm{X}}+\gamma_{\mathrm{C}}}{2},
    \label{EQN1}
\end{equation}

where $\Omega_{\mathrm{R}}$ is the observed Rabi splitting, and $\gamma_{\mathrm{X}}$ and $\gamma_{\mathrm{C}}$ are the exciton and cavity mode linewidths, respectively. The exciton linewidth extracted from D/WS$_2$/D PL is $\gamma_{\mathrm{X}}^{\mathrm{D}}\approx\SI{29.8}{\meV}$, while for the hBN-encapsulated reference we measure a room-temperature monolayer PL linewidth of $\gamma_{\mathrm{X}}^{\mathrm{hBN}}\approx\SI{22}{\meV}$. For the photonic contribution, we note that the linewidth $\gamma_{\mathrm{LP}}^{\mathrm{cav}}\approx\SI{9.5}{\meV}$ is extracted from the cavity-like lower polariton emission in the strong-coupling PL data at large negative detuning. This quantity is therefore not a direct measurement of the bare-cavity photon decay rate; instead it should be regarded as a conservative upper bound on the effective cavity linewidth inferred from emission in the coupled system (including additional broadening from the PL extraction and residual inhomogeneity). 

Even adopting the conservative estimate $\gamma_{\mathrm{C}}\equiv\gamma_{\mathrm{LP}}^{\mathrm{cav}}$, the observed Rabi splitting $\Omega_{\mathrm{R}}$ well exceeds the arithmetic mean of the linewidths as defined in Eq. \eqref{EQN1}, consistent with operation in the strong-coupling regime.

%remains comparable to and slightly larger than $\gamma_{\mathrm{X}}$ and clearly exceeds the polariton linewidths

Beyond the bare dispersions, we additionally analyze the \emph{momentum-dependent} PL intensity using a Boltzmann-population model evaluated at $T=\SI{300}{\kelvin}$, in which a thermal occupation of polariton states is weighted by the photonic Hopfield coefficient and convolved with a Lorentzian broadening (Supplementary Note~2) \cite{Lundt2016}. The resulting simulated spectra are shown in the left sides of Fig.~\ref{fig: planar-SC}a--f and reproduce the main trend of intensity redistribution along the lower polariton branch as detuning is varied. Notably, under nominally identical excitation and collection conditions, the D/WS$_2$/D data show closer agreement with this model than the hBN-encapsulated reference (quantified in Supplementary Note~3), consistent with reduced in-plane energy disorder. Such encapsulation-dependent differences may reflect variations in microscopic scattering pathways in WS$_2$ microcavities, where phonon-assisted and disorder-assisted processes can both contribute and their relative importance can depend on the local energy landscape and excitonic linewidth~\cite{An2026}. 

Taken together, the pronounced anti-crossing, the coupled-oscillator fits over a broad detuning range, and the consistency between measured and modelled momentum-dependent intensities provide a robust room-temperature benchmark of exciton--photon strong coupling in the open-cavity platform and establish D/WS$_2$/D as a highly suitable large-area active medium for subsequent polaritonic lattice experiments.

\begin{figure}[t]
\hspace*{\fill}
\includegraphics[width=1\textwidth]{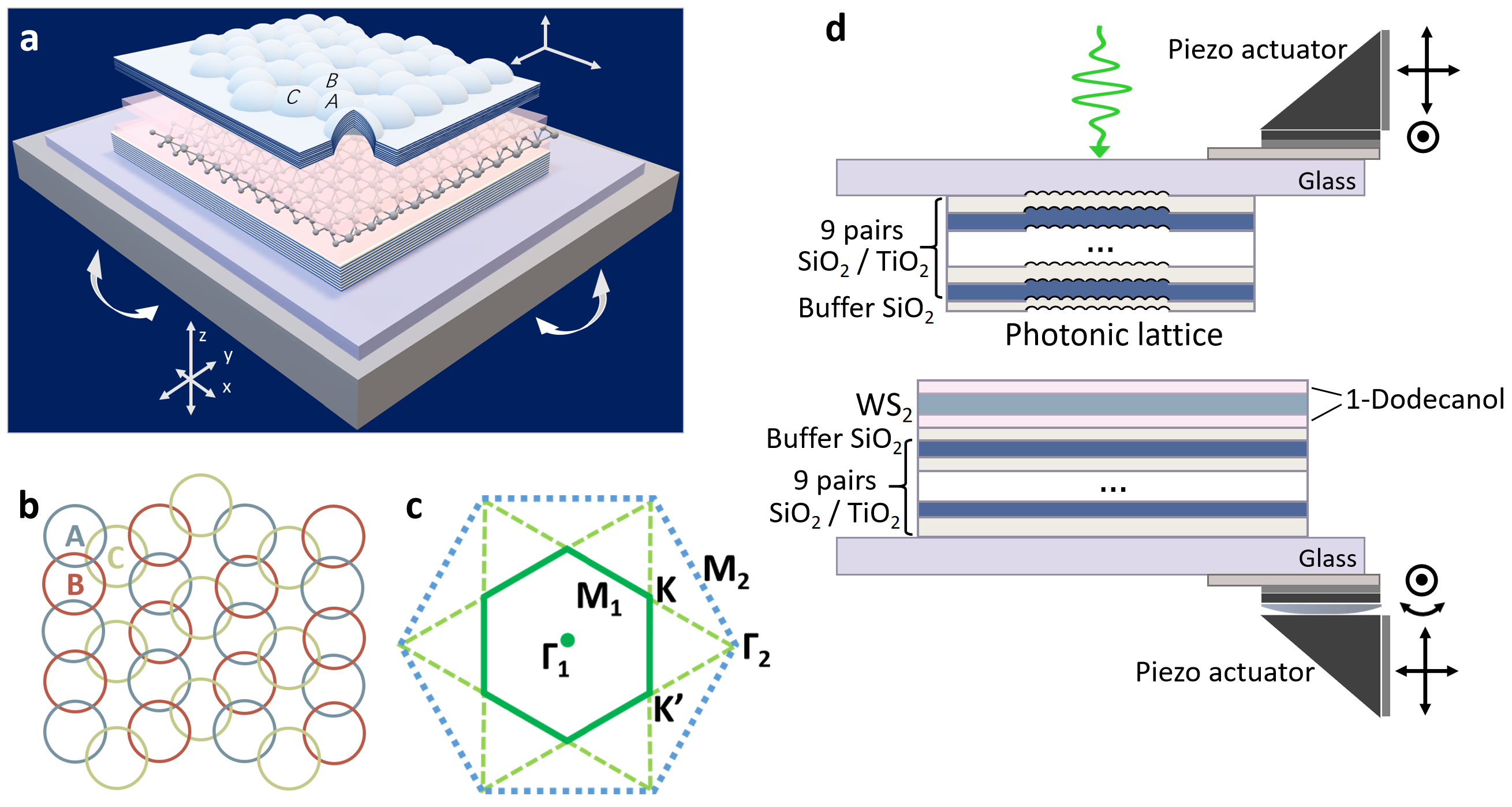}
\hspace*{\fill}
\caption{\textbf{Open-cavity implementation of a polaritonic kagome lattice.}
\textbf{a}, Schematic representation of an open microcavity in which the top mirror is patterned into a kagome photonic lattice and the bottom mirror hosts a D/WS$_2$/D monolayer.
\textbf{b}, Real-space kagome lattice, where A, B and C denote the three sites of the unit cell.
\textbf{c}, Geometry of the kagome lattice in reciprocal space, showing the high-symmetry points $\Gamma$, $K$ and $M$ and the first three Brillouin zones.
\textbf{d}, Cross-sectional schematic of the open cavity, illustrating the 1-dodecanol-encapsulated WS$_2$ monolayer on the bottom distributed Bragg reflector (DBR) and the top DBR mounted on piezoelectric nanopositioners that control cavity length, lateral alignment and tilt.}
\label{OC_Kagome_schematic}
\end{figure}

\subsection*{Exciton-polaritons in a 2D kagome lattice potential}

To demonstrate that our macroscopic polariton platform enables complex band-structure engineering, we realize an exciton–polariton kagome lattice as a proof of concept (Fig.~\ref{OC_Kagome_schematic}). The kagome photonic potential is realized by patterning the top mirror, whose periodic geometry is confirmed by AFM characterization (Supplementary Fig.~S3), while the bottom mirror hosts the D/WS$_2$/D monolayer (Fig.~\ref{OC_Kagome_schematic}a). The real-space lattice is based on a three-site unit cell (A, B and C sublattices; Fig.~\ref{OC_Kagome_schematic}b), and the corresponding reciprocal-space geometry with the high-symmetry points $\Gamma$, $K$ and $M$ is shown in Fig.~\ref{OC_Kagome_schematic}c. The cross-sectional cavity layout, including the D/WS$_2$/D monolayer on the bottom DBR and the nanopositioner-controlled top DBR, is shown in Fig.~\ref{OC_Kagome_schematic}d. 
%In the lowest band manifold, the Kagome geometry supports dispersive bands with Dirac-cone-like features and an additional weakly dispersive band arising from the lattice interference geometry.

\begin{figure}[t]
  \hspace*{\fill}
  \includegraphics[width=1\textwidth]{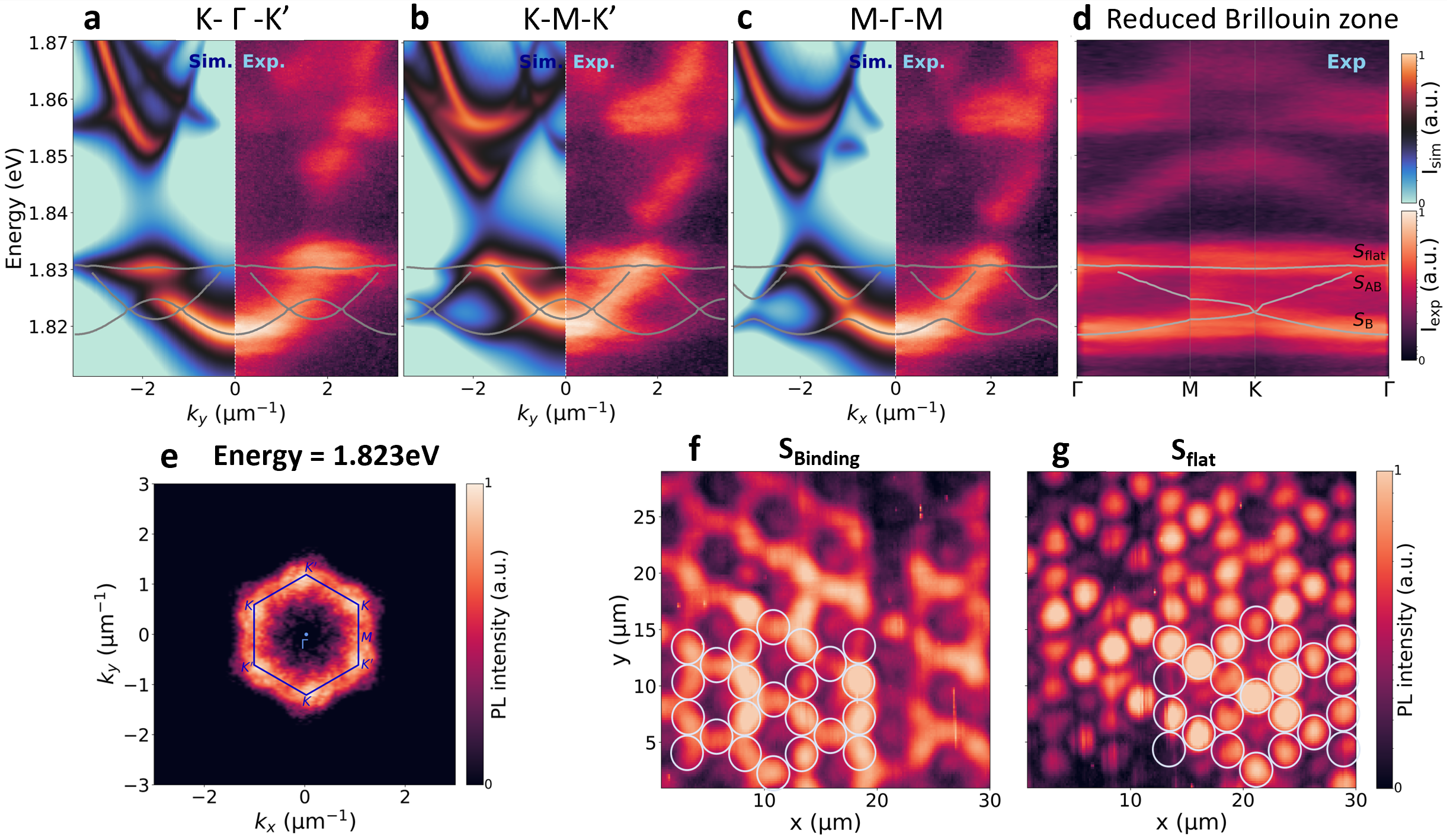}
  \hspace*{\fill}
  \caption{\textbf{Band structure and mode characterization of the WS$_2$ polaritonic kagome lattice.}
  \textbf{a}–\textbf{c}, Polariton dispersions along the high-symmetry directions $K$–$\Gamma$–$K^\prime$ (\textbf{a}), $K$–$M$–$K^\prime$ (\textbf{b}) and $M$–$\Gamma$–$M$ (\textbf{c}). For each cut, the left panel shows the calculated band structure, and the right panel displays the corresponding momentum-resolved PL, respectively. Gray solid lines overlaying the experimental spectra trace the calculated $s$-band peak positions.
  \textbf{d}, PL spectrum in a reduced Brillouin-zone representation, obtained by folding emission from higher Brillouin zones back into the first one. Gray lines indicate the calculated $s$-bands (binding, antibinding and flat band) from a linear Gross-Pitaevskii model.
  \textbf{e}, Fourier-space PL image at the energy of the Dirac points in the $s$-bands, highlighting the hexagonal first Brillouin zone and its high-symmetry points.
  \textbf{f}, \textbf{g}, Spectrally integrated real-space PL images at the energies of the binding $s$-band (\textbf{f}) and the flat band (\textbf{g}), revealing bond-centered emission for the binding band and site-localized emission for the flat band.
  The $k$-space data were processed by sixfold symmetrization, averaging over symmetry-related sectors of the Brillouin zone to improve the signal-to-noise ratio.}
  \label{fig: Kagome_results}
\end{figure}

In Fig.~\ref{fig: Kagome_results}, we characterize the band structure of exciton–polaritons in a WS$_2$ kagome lattice using angle-resolved PL tomography. The experimental dispersions (Fig.~\ref{fig: Kagome_results}a–c, right panels) reveal two low-energy dispersive subbands within the lowest $s$ band. They form characteristic Dirac cones at the $K$ and $K^\prime$ points, while a higher-energy subband is nearly flat and exhibits only weak dispersion. These features are visible along the high-symmetry directions $K$–$\Gamma$–$K^\prime$, $K$–$M$–$K^\prime$ and $M$–$\Gamma$–$M$ shown in Fig.~\ref{fig: Kagome_results}a–c.

The corresponding simulation results (Fig.~\ref{fig: Kagome_results}a–c, left panels) are obtained by solving a linearized Gross–Pitaevskii equation for the lower-polariton field in a kagome lattice potential~\cite{Carusotto2013_QuantumFluids}. Here, we neglect polariton–polariton interactions and pump–decay terms, so that the model describes the low-power, linear regime in the single-particle Schrödinger limit. The gray solid lines overlaying the experimental spectra trace the peak positions of the calculated $s$-bands, and show that the observed band structure is well captured by this linear, non-interacting model.

To enhance the visibility of weakly emitting modes, we employ a reduced-zone representation in which emission from higher Brillouin zones is folded back into the first Brillouin zone. The resulting spectrum in Fig.~\ref{fig: Kagome_results}d resolves the binding, antibinding and flat $s$-bands more clearly and provides a complete picture of the polariton band structure.

Further insight into the momentum-space emission is obtained from the Fourier-space PL image in Fig.~\ref{fig: Kagome_results}e, recorded at the Dirac-point energy within the lowest $s$-band manifold. The emission intensity is concentrated near the six vertices of the hexagonal first Brillouin zone, corresponding to the $K$ and $K^\prime$ points. This observation is consistent with the presence of Dirac-point-like features in the dispersive $s$-bands and indicates that, at this energy, the detected emission is dominated by polariton states at momenta close to $K/K^\prime$.

Energy-resolved real-space imaging in Fig.~\ref{fig: Kagome_results}f,g reveals distinct spatial emission patterns for different states within the lowest $s$-band manifold. At the binding $s$-band energy (Fig.~\ref{fig: Kagome_results}f), the emission intensity is enhanced on the links between neighbouring lattice sites, giving a bond-centered pattern. In contrast, at the energy of the weakly dispersive $s$-band (Fig.~\ref{fig: Kagome_results}g), the emission is predominantly concentrated on the lattice sites, yielding a site-centered pattern. These complementary real-space signatures are consistent with the different mode characters expected from the kagome lattice geometry and provide an intuitive visualization of the bonding and flat-band-like states.

% =========================================================
\section*{Conclusion}
% =========================================================

We demonstrate a tunable open-microcavity platform based on macroscopic, optically uniform WS$_2$ monolayers that realizes room-temperature exciton--polaritons, with a polaritonic kagome lattice serving as a proof of concept for complex band-structure engineering. Using gold-tape-assisted exfoliation together with 1-dodecanol encapsulation, we obtain millimeter-scale monolayers with homogeneous optical properties over lateral length scales of order \SI{300}{\micro\meter}, providing a reproducible active medium for cavity experiments.

In a planar open cavity, we benchmark strong coupling under identical measurement conditions for an hBN-encapsulated WS$_2$ reference flake and for the large-area D/WS$_2$/D monolayer. In both cases, we observe well-defined upper and lower polariton branches with a pronounced anti-crossing as the cavity mode is tuned through the exciton. For the D/WS$_2$/D sample we extract a Rabi splitting of $\hbar\Omega_{\mathrm{R}}=\SI[separate-uncertainty]{31 \pm 0.5}{\meV}$, consistent with the strong-coupling regime. A Boltzmann-population model evaluated at $T=\SI{300}{\kelvin}$ reproduces the momentum-dependent redistribution of PL intensity and shows closer agreement for D/WS$_2$/D than for the hBN reference, consistent with reduced in-plane disorder and improved uniformity in the macroscopic monolayer.

We implement a polaritonic kagome lattice by patterning the top mirror. Angle-resolved tomography reveals the characteristic kagome $s$-band manifold with Dirac-cone-like dispersive bands and a nearly dispersionless band, in good agreement with a linear, non-interacting model. Complementary real-space imaging resolves distinct mode patterns at the energies of the dispersive and nearly flat bands, corroborating the lattice-mode assignment.

Together, these results position macroscopic TMDC monolayers in open microcavities as a large-area and tunable basis for room-temperature polariton lattices in engineered two-dimensional geometries. The combination of \textit{in situ} detuning control and patterned photonic potentials opens routes towards systematic studies of polariton band engineering and driven--dissipative many-body effects, and can be extended to more complex TMDC heterostructures, including twisted and moir\'e-aligned bilayers with additional internal degrees of freedom~\cite{Zhang2021_MoirePolaritons, Han2025Infrared, Zhao2025Room}. Looking ahead, this platform provides a versatile testbed for exploring emergent quantum and topological phenomena in nonequilibrium bosonic systems under ambient conditions. Its scalability and design flexibility further make it a promising foundation for polaritonic simulators and low-power photonic technologies based on engineered light–-matter states \cite{Huber2026,Zacheo2026}.

\bibliographystyle{sn-nature}
\bibliography{references}

% =========================================================
\section*{Methods}
% =========================================================

\subsection*{Mirror fabrication}
The bottom distributed Bragg reflector (DBR) was fabricated on a glass substrate that was first patterned into a mesa structure by HF wet etching. The resulting mesa height is approximately \SI{80}{\micro\meter}, which defines the cavity region. A dielectric DBR consisting of nine pairs of $\lambda/4$-thick TiO$_2$/SiO$_2$ layers, designed for a center wavelength of $\lambda_c = \SI{620}{nm}$, was then sputter-deposited onto the entire substrate by over sputtering in a Nordiko 3000 system. An additional SiO$_2$ spacer layer was deposited on top of the DBR to act as a buffer layer, so that emitting materials such as the WS$_2$ monolayer are positioned close to an antinode of the cavity field once the cavity is assembled. The bottom cavity mesa has a lateral area of about \SI{1}{\square\milli\meter}.

The top DBR was fabricated in the same sputtering system on a separate substrate and also consists of nine TiO$_2$/SiO$_2$ quarter-wave pairs centered at $\lambda_c = \SI{620}{nm}$. A mesa structure with a height of approximately \SI{80}{\micro\meter} and a lateral area of about $\SI{300}{\micro\meter}\times\SI{300}{\micro\meter}$ defines the active cavity region on the top mirror. Its surface was patterned by focused ion beam milling in a FEI Helios Dual Beam system. The kagome lattice was defined using Ga$^+$ ions with an emission current of \SI{7}{\nano\ampere}, an acceleration voltage of \SI{30}{\kilo\volt}, a dwell time of \SI{15}{\micro\second} and 20 passes.

\subsection*{Gold-tape exfoliation and 1-dodecanol encapsulation}
D/WS$_2$/D monolayers were fabricated on distributed Bragg reflector (DBR) substrates in a cleanroom environment. Large-area monolayer WS$_2$ was prepared by ultraflat Au-assisted exfoliation based on template stripping, following previously reported procedures~\cite{Science_Gold-tape-exfoliation2020}. Briefly, a \SI{150}{nm}-thick Au film was deposited on a Si wafer and stripped using thermal-release tape with a polyvinylpyrrolidone (PVP) interfacial layer to expose an ultraflat Au surface. A bulk WS$_2$ crystal was brought into contact with this Au surface for \SI{1}{min}, after which monolayer WS$_2$ was exfoliated onto the Au-coated tape.

Before transfer, the bottom 1-dodecanol layer was formed on the DBR substrate by drop-casting 1-dodecanol while the substrate was held at \SI{180}{\celsius}, followed by heating for \SI{2}{min}~\cite{Li_Dodecanol-WS2_2023}. Residual molecules were removed by immersion in isopropanol, and the substrate was dried with nitrogen gun. The Au-supported WS$_2$ monolayer was then aligned and transferred onto the 1-dodecanol-functionalized DBR substrate under an optical microscope with a three-axis alignment stage. After alignment and contact with the target 1-dodecanol-functionalized DBR substrate, the thermal-release tape was detached. The PVP interfacial layer was subsequently dissolved, and the Au layer was removed using a mild I$_2$/I$^{-}$ etchant solution. The 1-dodecanol functionalization was then repeated on top of the transferred WS$_2$ monolayer to complete the D/WS$_2$/D structure. Additional fabrication details are provided in Supplementary Fig.~S1 and Supplementary Note~1.

\subsection*{Optical measurements}
PL measurements were performed in a home-built Fourier-imaging microscope using a continuous-wave laser at \SI{532}{nm}. The excitation beam was focused onto the sample through an $f = \SI{300}{mm}$ lens and a high-numerical-aperture objective (NA = 0.65, 50$\times$). The emitted (or reflected) signal was collected by the same objective and passed through a long-pass filter to suppress the excitation light. A Fourier-space image plane of the objective back aperture was formed and relayed in a 4$f$ configuration to generate a conjugate real-space image plane, where a confocal pinhole was used to spatially select the region of interest.

The transmitted signal was then imaged onto an Andor Newton 971 EM-CCD camera using an $f = \SI{150}{mm}$ collimating lens and an $f = \SI{300}{mm}$ imaging lens for $k$-space or real-space PL imaging. For angle-resolved spectroscopy, the Fourier plane was re-imaged onto the entrance slit of a spectrometer equipped with three interchangeable gratings (150, 300 and 1200 lines\,mm$^{-1}$) and a motorized entrance slit.

% =========================================================
\section*{Author contributions}
% =========================================================

S.Hu. and S.K. conceived the idea. J.Q., E.W. and L.M. fabricated and optimized the encapsulated monolayer material. J.D., D.H., M.E. and M.K. fabricated the patterned and unpatterned optical mirrors. S.Hu., S.S., J.D., D.H. and S.B. build up the open-cavity setup and performed the spectroscopic measurements. S.Hu. and S.S. analyzed the data. S.Hu., J.D., and S.W. carried out the simulations. S. Hu. and S.K. drafted the manuscript and all authors contributed to editing and review. A.C., S.Hö. and S.K. provided funding and supervised the work.

% =========================================================
\section*{Funding}
% =========================================================

The authors acknowledge financial support by the DFG via the Würzburg-Dresden Cluster of Excellence on Complexity and Topology in Quantum Matter (ctd.qmat) (EXC 2147, Project ID 390858490).

% =========================================================
\section*{Competing interests}
% =========================================================

The authors declare no competing interests.

% =========================================================
% Bibliography: include this only once
% =========================================================

% =========================================================
% Supplementary material
% =========================================================

\clearpage
\setcounter{page}{1}
\renewcommand{\thepage}{S\arabic{page}}

%\section*{Supplementary Information}

%\clearpage
\appendix

\begin{center}
    {\LARGE\bfseries Supplementary Information}\\[1.5em]
    {\large Encapsulated macroscopic WS$_2$ monolayers enable room-temperature exciton-polariton lattices}
\end{center}

\renewcommand{\thesection}{S\arabic{section}}
\renewcommand{\thesubsection}{S\arabic{section}.\arabic{subsection}}
\renewcommand{\thefigure}{S\arabic{figure}}
\renewcommand{\thetable}{S\arabic{table}}
\renewcommand{\theequation}{S\arabic{equation}}

\setcounter{section}{0}
\setcounter{figure}{0}
\setcounter{table}{0}
\setcounter{equation}{0}

\begin{figure}[h]
\centering
\includegraphics[width=\textwidth]{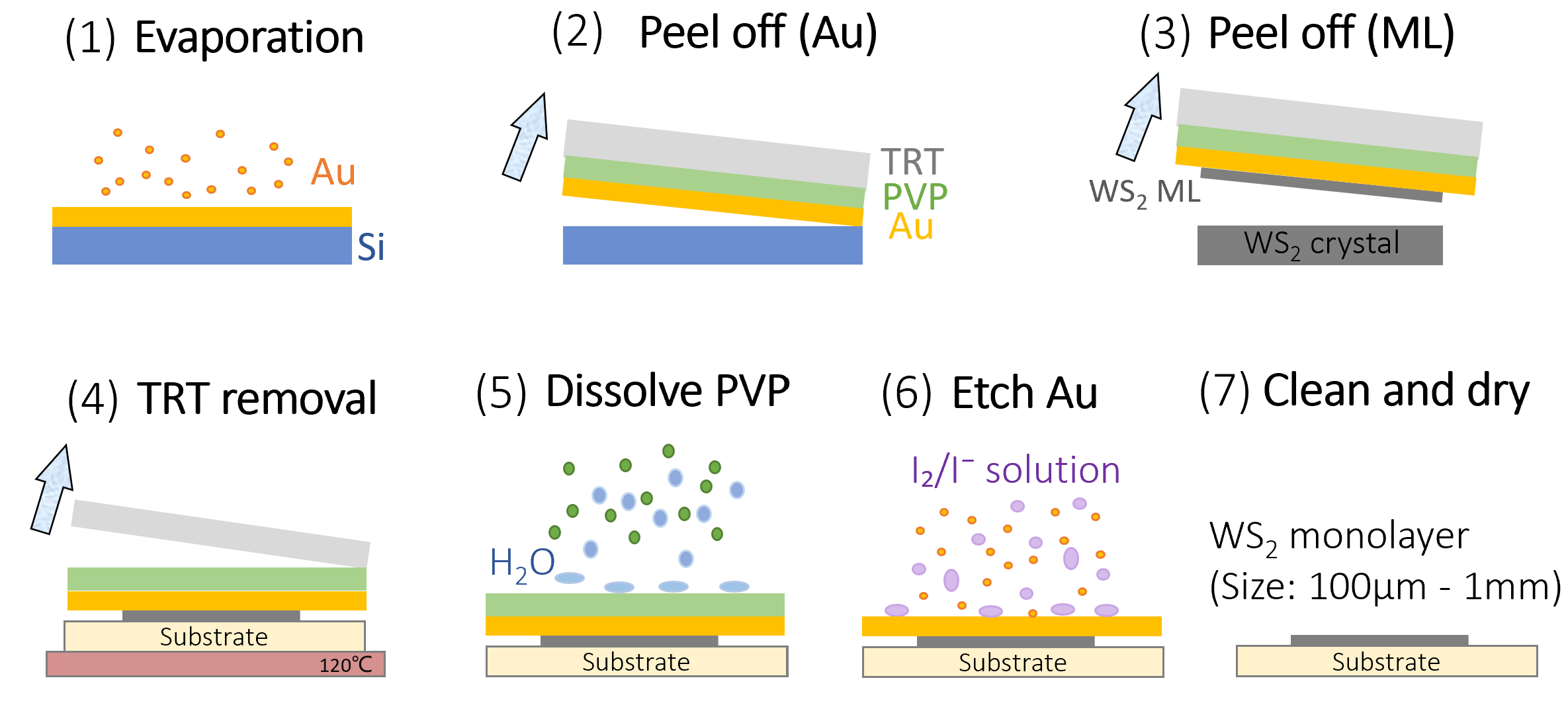}
\caption{\textbf{Schematic of the gold-tape-assisted exfoliation and transfer of WS$_2$ monolayers.}
Fabrication steps following the gold-tape method of Ref.~\cite{Science_Gold-tape-exfoliation2020}. 
(1) A thin gold (Au) film is deposited on a Si wafer to provide an atomically flat metal surface.  
(2) The Au layer is released from the Si wafer using a polyvinylpyrrolidone (PVP) interlayer and thermal release tape (TRT), forming a flexible Au/PVP/TRT stack. 
(3) The Au/PVP/TRT stack is brought into contact with bulk WS$_2$, where strong S--Au affinity leads to selective pickup of the top WS$_2$ monolayer (ML) when the stack is peeled off from the crystal. 
(4) The Au/PVP/WS$_2$ stack on TRT is brought into contact with the target substrate (bottom distributed Bragg reflector) and heated to about \SI{120}{\celsius} to release the TRT. 
(5) The PVP layer is dissolved in water, leaving the Au/WS$_2$ stack on the substrate. 
(6) The Au film is removed by wet etching in an I$_2$/I$^-$ solution, exposing the WS$_2$ monolayer on the substrate. 
(7) After rinsing and drying, a clean WS$_2$ monolayer with a typical lateral size of \SIrange{100}{1000}{\micro\meter} remains on the cavity substrate.
Further details are provided in Supplementary Note~1.}
\label{fig:SI_gold_tape}
\end{figure}

\clearpage

%sample_D-hBN
\begin{figure}[h]
\centering
\includegraphics[width=\textwidth]{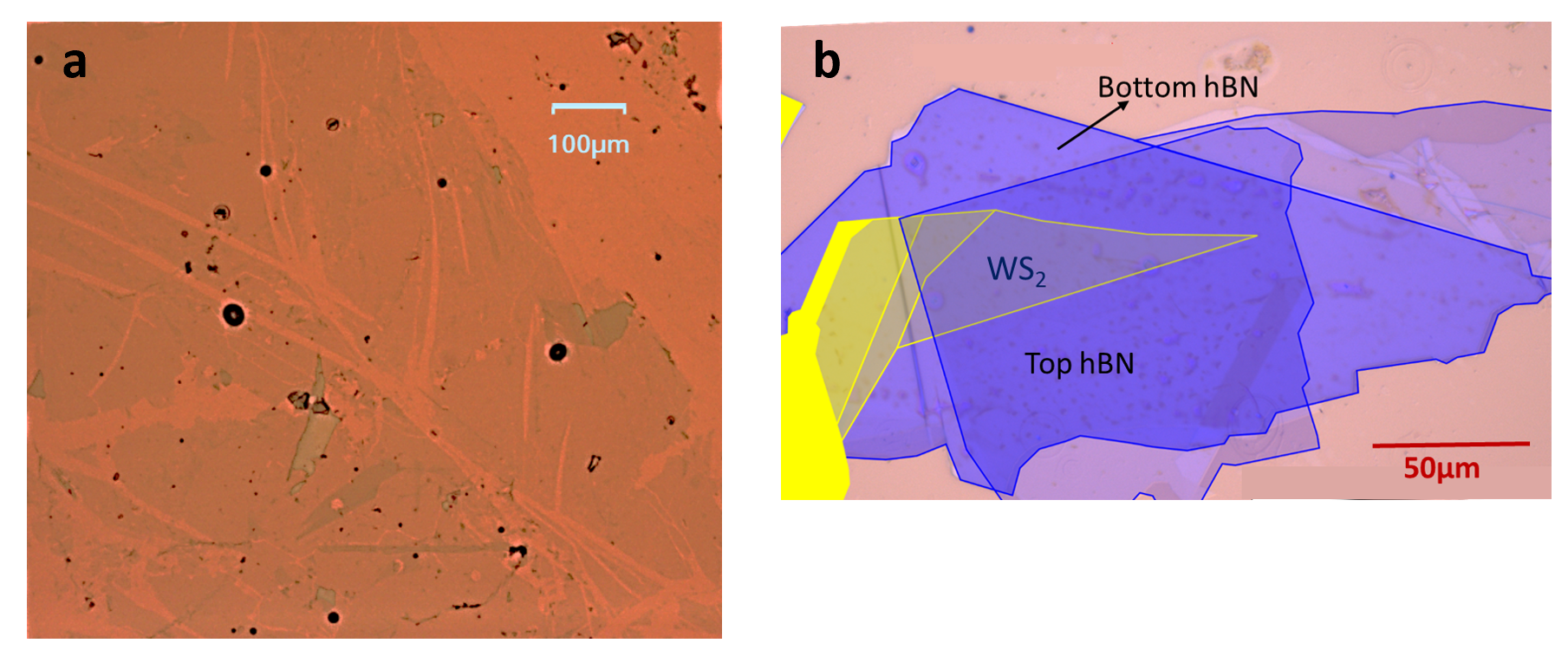}
\caption{\textbf{Optical microscope images of WS$_2$ samples on DBR substrates.}
\textbf{a}, Optical microscope image of the 1-dodecanol-encapsulated macroscopic WS$_2$ monolayer (D/WS$_2$/D) on a distributed Bragg reflector (DBR) substrate. The large, optically uniform region corresponds to the encapsulated WS$_2$ monolayer, whereas the higher-contrast patches indicate thicker WS$_2$ regions (multilayer or bulk). The scale bar is \SI{100}{\micro\meter}. 
\textbf{b}, Optical microscope image of the hBN-encapsulated WS$_2$ reference sample (hBN/WS$_2$/hBN) on a DBR substrate. The outlines of the top and bottom hBN flakes and the WS$_2$ monolayer region are indicated. The bright/yellow region corresponds to bulk WS$_2$. The scale bar is \SI{50}{\micro\meter}.}
\label{fig:sample_D-hBN}
\end{figure}

\clearpage

%\section*{Kagome lattice}

\begin{figure}[h]
\centering
  \includegraphics[width=0.92\textwidth]{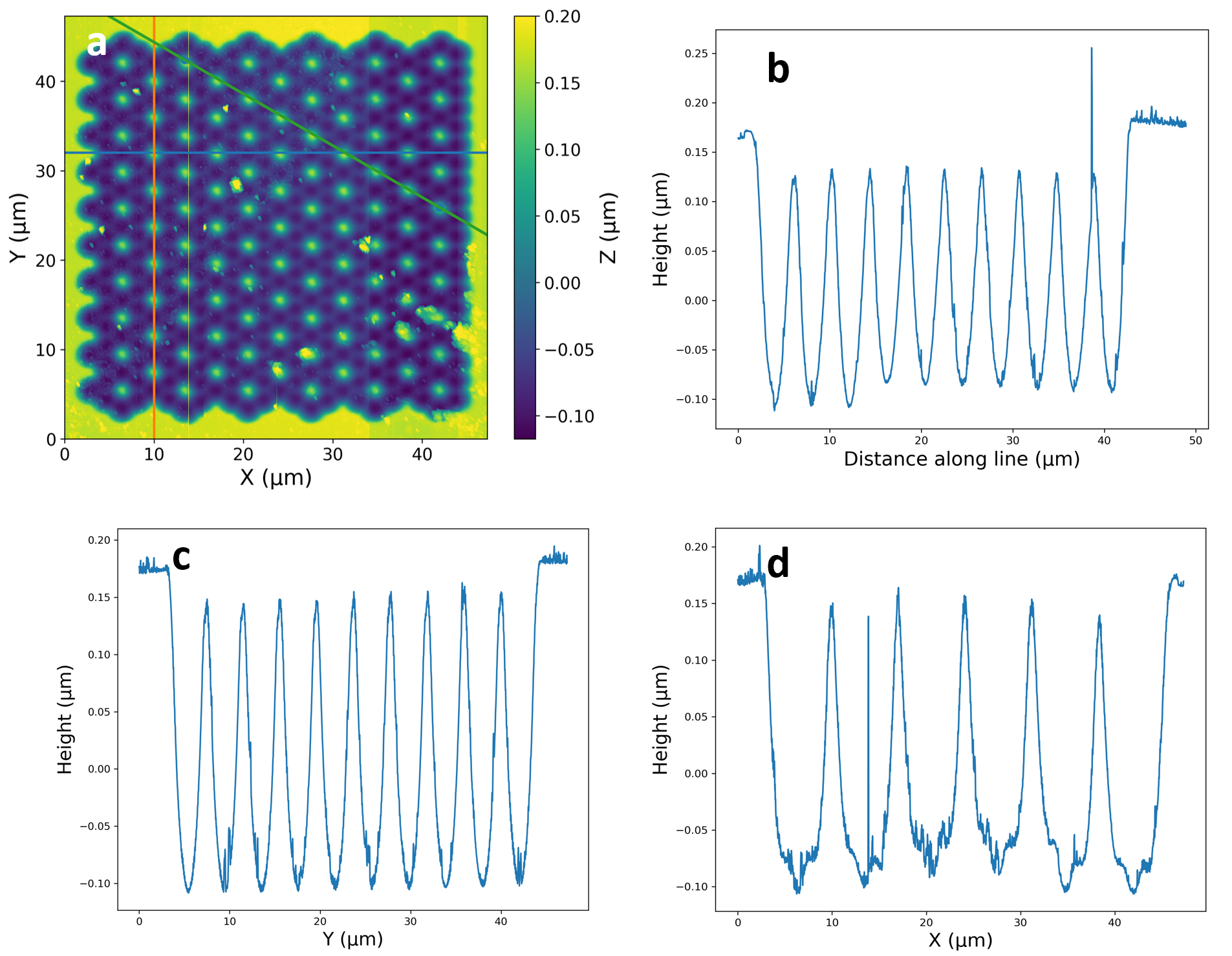} 
  \caption{\textbf{AFM characterization of the patterned Kagome lattice on the top mirror.}
  \textbf{a}, Atomic force microscopy (AFM) height map of the Kagome lattice patterned into the top DBR mirror (scan area $\sim$\,\SI{40}{\micro\meter}$\times$\SI{40}{\micro\meter}). The lattice consists of circular traps with a designed diameter $d=\SI{4}{\micro\meter}$ arranged in a Kagome geometry with center-to-center distance $a$, corresponding to a reduced trap distance $v=a/d=0.8$. The colored lines indicate representative line cuts through the AFM map: the green line marks a diagonal cut, the blue line marks a horizontal cut, and the orange line marks a vertical cut. 
  \textbf{b}, Height profile along the diagonal cut (green line in \textbf{a}). 
  \textbf{c}, Height profile along the vertical cut (orange line in \textbf{a}). 
  \textbf{d}, Height profile along the horizontal cut (blue line in \textbf{a}). 
  Together, the 2D height map and line profiles confirm the periodicity and uniformity of the patterned Kagome lattice and provide a direct characterization of the trap geometry used for the polaritonic Kagome lattice measurements.}
  \label{fig: Kagome_AFM} 
\end{figure}
 
\clearpage

%\section*{Cavity design}

%\textcolor{green}{Q factor from PL!!!}

\begin{figure}[htbp]
\centering
\includegraphics[width=\textwidth]{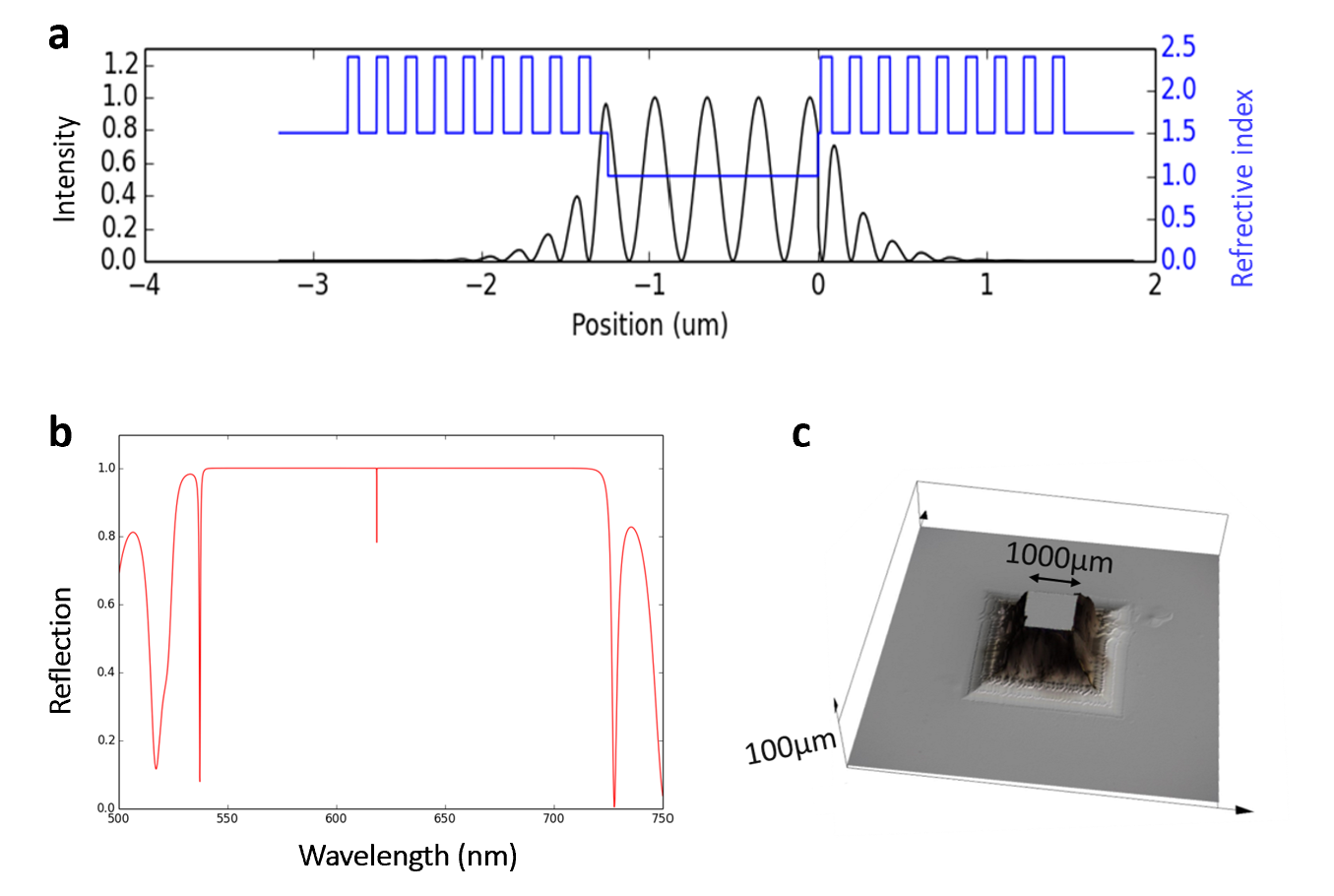}
\caption{\textbf{Design of the planar cavity and mesa geometry.}
\textbf{a}, Simulated refractive-index profile of the distributed Bragg reflector (blue) and corresponding longitudinal intensity profile of the fundamental cavity mode (black) as a function of position along the growth direction. The mode is centered in the air gap between the two DBRs and exhibits an antinode at the position of the WS$_2$ monolayer on the bottom mirror.
\textbf{b}, Transfer-matrix calculation of the reflectivity spectrum of the planar cavity, showing a high-reflectivity stop band around \SI{620}{\nano\meter} with a narrow cavity resonance inside the stop band.
\textbf{c}, Three-dimensional profilometry image of the etched mesa on the bottom substrate, with a lateral size of approximately \SI{1000}{\micro\meter} and a height of about \SI{100}{\micro\meter}, which defines the active cavity region.}
\label{fig:SI_cavity_design}
\end{figure}

%\clearpage

\begin{figure}[htbp]
\centering
\includegraphics[width=1\textwidth]{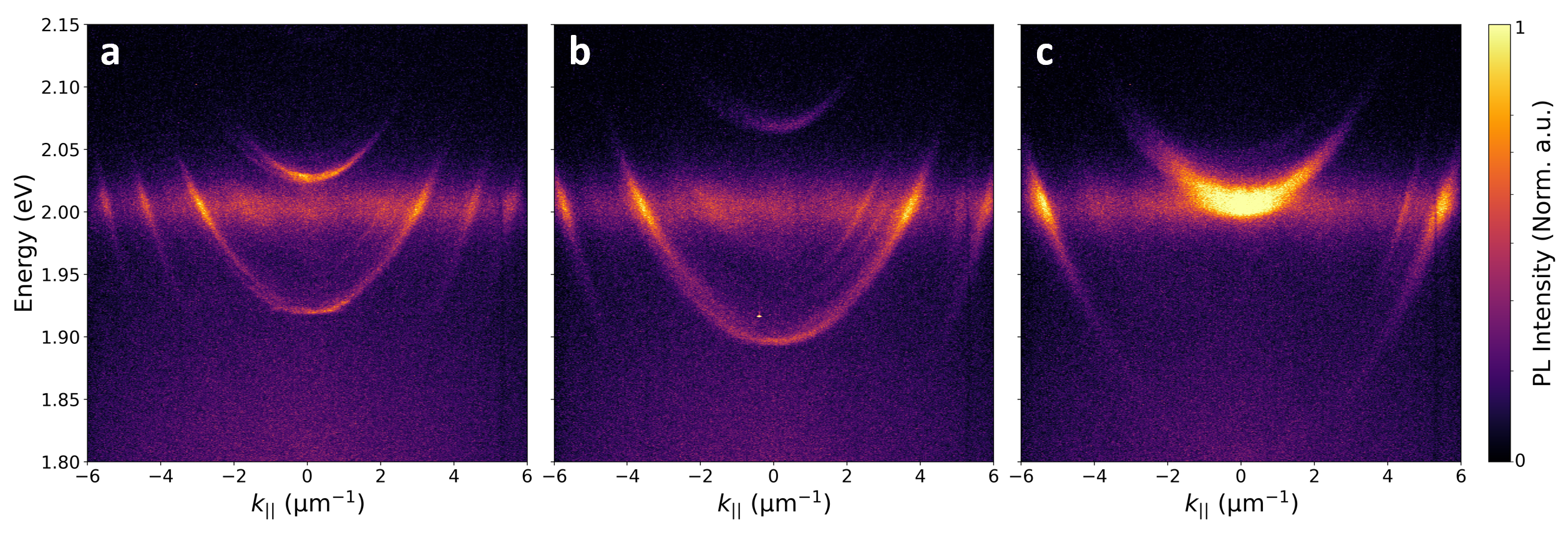}
\caption{\textbf{Longitudinal cavity modes versus cavity length.} Momentum-resolved photoluminescence spectra of the bare open cavity for different mirror separations $L_{\mathrm{cav}}$.
The cavity lengths are \textbf{a}, $L_{\mathrm{cav}} \approx \SI{5.8}{\micro\metre}$, \textbf{b},
$L_{\mathrm{cav}} \approx \SI{3.6}{\micro\metre}$ and \textbf{c},
$L_{\mathrm{cav}} \lesssim \SI{1.6}{\micro\metre}$, as estimated from the free spectral range of the longitudinal cavity modes.
For the longest cavity (\textbf{a}) several parabolic photonic branches with small energy spacing are visible.
As $L_{\mathrm{cav}}$ is reduced (\textbf{b}), the spacing between successive cavity modes increases and fewer branches appear within the same spectral range.
For the shortest separation (\textbf{c}), the mode spacing is largest and only a single cavity branch remains in the detection window.
All spectra are measured at room temperature under CW \SI{532}{\nano\metre} excitation, and the color scale indicates the normalized PL intensity.
}
\label{fig:SI_detuning}
\end{figure}

\begin{figure}[htbp]
\centering
\includegraphics[width=0.7\textwidth]{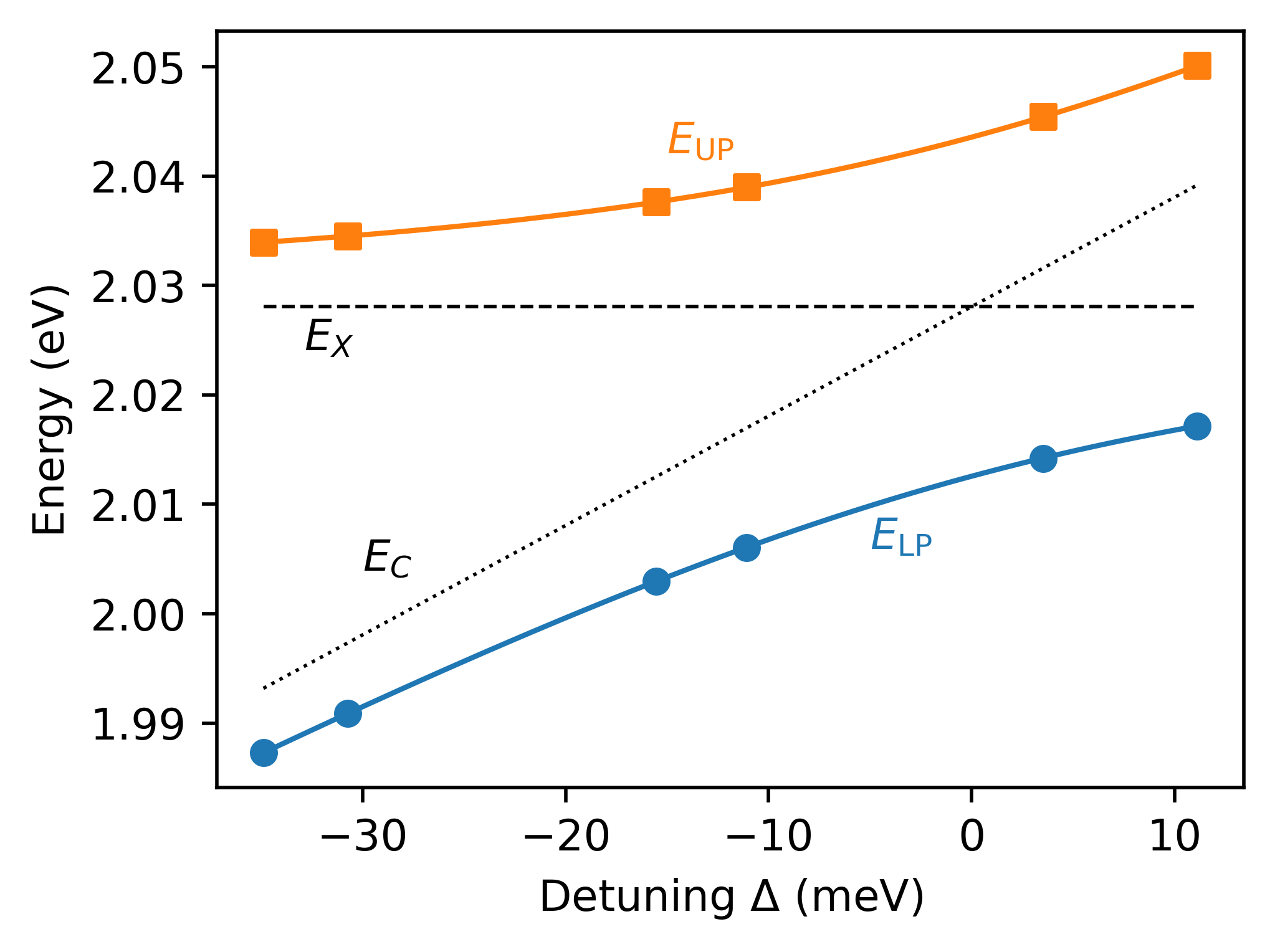}
\caption{\textbf{Detuning dependence of the polariton energies.}
Energies of the lower (E$_\mathrm{LP}$, blue circles) and upper (E$_\mathrm{UP}$, orange squares) polariton branches as a function of exciton–cavity detuning $\Delta$, extracted from angle-resolved PL spectra.
Solid lines show the corresponding fits of a coupled-oscillator model.
The dashed horizontal line marks the bare exciton energy E$_\mathrm{X}$, and the dotted line indicates the uncoupled cavity mode $E_{\mathrm{C}} = E_{\mathrm{X}} + \Delta$.
The avoided crossing between E$_\mathrm{LP}$ and E$_\mathrm{UP}$ as $\Delta$ is tuned through zero illustrates the strong-coupling regime of the D/WS$_2$/D monolayer in the open cavity.}
\label{fig:SI_anticrossing}
\end{figure}

\section*{Supplementary Note 1: Sample fabrication}
The D/WS$_2$/D samples were fabricated on DBR substrates using a sequential process combining ultraflat Au-assisted exfoliation, transfer onto a bottom 1-dodecanol interfacial layer and subsequent top 1-dodecanol encapsulation. All fabrication steps were carried out in a cleanroom environment to minimize contamination.

For Au-assisted exfoliation, a \SI{150}{nm}-thick Au film was deposited on a Si wafer and stripped from the substrate using thermal-release tape (TRT) with a polyvinylpyrrolidone (PVP) interfacial layer, exposing an ultraflat Au surface. A bulk WS$_2$ crystal was brought into contact with the Au surface for \SI{1}{min}. The Au/PVP/TRT stack was then peeled from the crystal, selectively picking up the top WS$_2$ monolayer, as schematically shown in Supplementary Fig.~S1.

Before monolayer transfer, the bottom 1-dodecanol layer was prepared on the DBR substrate. 1-Dodecanol was drop-cast onto the DBR while the substrate was held at \SI{180}{\celsius} on a hot plate, followed by heating for \SI{2}{min}. Excess 1-dodecanol was removed by gently immersing the substrate in isopropanol, and the substrate was dried with nitrogen.

The Au-supported WS$_2$ monolayer was aligned with the DBR mesa under an optical microscope equipped with a three-axis alignment stage and brought into contact with the 1-dodecanol-functionalized substrate. The TRT was released, the PVP interfacial layer was dissolved in water, and the Au layer was removed using a mild I$_2$/I$^{-}$ etchant solution. Finally, the same 1-dodecanol functionalization procedure was repeated on top of the transferred WS$_2$ monolayer, completing the D/WS$_2$/D encapsulation.

\section*{Supplementary Note 2: Intensity simulation}
To model the momentum-resolved PL intensity, we assume that polaritons relax into a thermal population within each branch and that the detected signal is weighted by the photonic Hopfield coefficient \cite{Lundt2016}. We therefore write the simulated spectrum as
\begin{equation}
I_{\mathrm{PL}}^{\mathrm{model}}(E,k)\ \propto\ \sum_{j\in\{\mathrm{LP},\mathrm{UP}\}} |C_j(k)|^2\, \exp\!\left[-\frac{E_j(k)-E_0}{k_{\mathrm{B}}T}\right]\,
\frac{\gamma_j/2}{\bigl(E-E_j(k)\bigr)^2+(\gamma_j/2)^2},
\end{equation}
where $E_j(k)$ and $|C_j(k)|^2$ are obtained from the coupled-oscillator model, $T=\SI{300}{\kelvin}$, and $\gamma_j$ accounts for spectral broadening (Lorentzian). The overall prefactor and $E_0$ set the normalization and reference energy and do not affect the extracted dispersion. 

\section*{Supplementary Note 3: Model--experiment comparison and goodness-of-fit}

To quantify the agreement between the Boltzmann-population simulations and the measured momentum-resolved PL spectra, we evaluate the normalized root-mean-square error (NRMSE, normalized to the peak intensity of the corresponding experimental spectrum) and the coefficient of determination ($R^2$). Because the emission is dominated by the lower-polariton branch at low in-plane momenta, we focus on the low-$k_{\parallel}$ window ($0 \le k_{\parallel} \le \SI{1.0}{\micro\meter^{-1}}$), which contains the strongest signal and is most relevant for assessing the redistribution of population along the lower polariton.

\begin{table}[htbp]
\centering
\caption{\textbf{Goodness-of-fit metrics in the low-momentum window.} Metrics are computed between the simulated and experimental momentum-resolved PL spectra in the low-$k_{\parallel}$ region ($0 \le k_{\parallel} \le \SI{1.0}{\micro\meter^{-1}}$), where the lower-polariton emission is strongest. NRMSE is normalized to the experimental peak intensity. Values are given as mean $\pm$ s.e.m.\ across the detuning conditions analyzed.}
\label{tab:boltzmann_gof_lowk}
\begin{tabular}{lcc}
\hline
Sample & NRMSE/peak & $R^2$ \\
\hline
D/WS$_2$/D & $0.081 \pm 0.003$ & $0.857 \pm 0.012$ \\
hBN/WS$_2$/hBN & $0.109 \pm 0.004$ & $0.733 \pm 0.018$ \\
\hline
\end{tabular}
\end{table}

The statistics in Table~\ref{tab:boltzmann_gof_lowk} show systematically lower NRMSE and higher $R^2$ for the D/WS$_2$/D device compared with the hBN-encapsulated reference, indicating closer agreement with the thermal-population model in the momentum range that dominates the detected emission.

We emphasize that the Boltzmann-population model used here is intentionally minimal and captures only (i) the dispersion extracted from the coupled-oscillator model, (ii) a thermal occupation factor, (iii) photonic weighting via the Hopfield coefficient, and (iv) a phenomenological Lorentzian broadening. At larger momenta, the measured intensity is typically weaker and more sensitive to experimental factors (finite collection aperture, background subtraction, and momentum-dependent sensitivity), and it can additionally be influenced by non-thermal relaxation pathways, scattering via disorder, and momentum-dependent radiative outcoupling. Consequently, deviations between model and experiment are expected to be more pronounced at high $k_{\parallel}$, and we do not interpret the model as a quantitative description of the full $k_{\parallel}$ window. Instead, the goodness-of-fit analysis is used as a compact, comparative metric to assess how reproducibly the low-$k_{\parallel}$ population redistribution is captured under identical measurement conditions for the two encapsulation strategies.

% Do not add another bibliography here.

\end{document}